\documentclass[10pt,twoside]{article}
\usepackage{graphicx}
\usepackage{amsmath,amsfonts,amscd}
\usepackage{Latex-document}

\newcommand{\adH}{\delta_H}

\newtheorem{theorem}{Theorem}[section]          %%[chapter]

\newcommand{\eq}[1]{(\ref{#1})}

\newcommand{\Rl}{\mathbb{R}}

\renewcommand{\vec}[1]{\boldsymbol{#1}}
\newcommand{\SKIP}[1]{\text{ [text skipped] } \newline}

\newcommand{\RR}{\mathbb R}

\newcommand{\be}{\begin{equation}}
\newcommand{\ee}{\end{equation}}
\newcommand{\bea}{\begin{eqnarray}}
\newcommand{\eea}{\end{eqnarray}}
\newcommand{\bean}{\begin{eqnarray*}}
\newcommand{\eean}{\end{eqnarray*}}
\newcommand{\beq}{\begin{equation}}
\newcommand{\eneq}{\end{equation}}
\newcommand{\bey}{\begin{eqnarray}}
\newcommand{\eey}{\end{eqnarray}}
\newcommand{\beyn}{\begin{eqnarray*}}
\newcommand{\eeyn}{\end{eqnarray*}}

\newcommand{\beann}{\begin{eqnarray*}}
\newcommand{\eeann}{\end{eqnarray*}}

\newcommand{\g}{\gamma}
\newcommand{\e}{\varepsilon}

\newcommand{\Tr}{\mbox{Tr}}

\newcommand{\ran}{\rangle}
\newcommand{\lan}{\langle}

\newcommand{\rcite}{\cite}

\newcommand{\fq}{\mathfrak{q}}

\newcommand{\fv}{{v}}

\newcommand{\bq}{{\bf q}}

\newcommand{\bu}{{\bf u}}

\title{\bf  Derivation of the Euler Equations\vskip -2mm
from Many-body Quantum Mechanics\thanks{Copyright \copyright\ 2002
Nachtergaele and Yau. Reproduction of this article in its
entirety, by any means, is permitted for non-commercial
purposes.}\vskip 6mm}
\author{{\bf Bruno Nachtergaele}\thanks{Department of Mathematics, University of California, Davis, CA 95616, USA.
E-mail: bxn@math.ucdavis.edu} \quad Horng-Tzer Yau\thanks{Courant
Institute, New York University, New York, NY 10012, USA. E-mail:
yau@cims.nyu.edu}\vspace*{-0.5cm}}
\date{\vspace{-8mm}}

\begin{document}
\maketitle

\begin{center}
{\it $($Dedicated to Andr\'e Verbeure on the occasion of his sixtieth birthday$)$}
\end{center}

\thispagestyle{first} \setcounter{page}{467}

\begin{abstract}\vskip 3mm
The Heisenberg dynamics of the energy, momentum, and particle densities for
fermions with short-range pair interactions is shown to converge to the
compressible Euler equations in the  hydrodynamic limit. The pressure function
is given by the standard formula from quantum statistical mechanics with the
two-body potential under consideration. Our derivation is based on a quantum
version of the entropy  method and a suitable quantum virial theorem. No
intermediate description, such as a Boltzmann equation or semi-classical
approximation, is used in our proof. We require some technical conditions on
the dynamics, which can be considered as interesting open problems in their
own right.
\vskip 4.5mm


\end{abstract}

\vskip 12mm

\section{Introduction} \label{sec:intro}\setzero
\vskip-5mm \hspace{5mm}

The fundamental laws of non-relativistic microscopic physics are {\em Newton's}
and the {\em Schr\"odinger} equation in the classical and the quantum case
respectively. These equations are impossible to solve for large systems and
macroscopic dynamics is therefore modeled by phenomenological equations such as
the {\em  Euler} or the {\em Navier-Stokes} equations. Although the latter were
derived centuries ago from  continuum  considerations, they are in principle
consequences  of  the microscopic physical laws and should be viewed as
secondary equations. It was first observed by Morrey \rcite{Mor} in the
fifties that the Euler equations become `exact' in the Euler limit, provided
that the solutions to the  Newton's equation  are `locally' in equilibrium.
Morrey's original work was far from rigorous and, in particular, the meaning of
`local equilibrium' was not clear.  It is nevertheless a very original work
which led to the later development of the hydrodynamical limits of interacting
particle systems. In terms of a rigorous proof along the lines of Morrey's
original argument, however, there has been little progress until the recent
work \cite{OVY}. This long delay is mostly due to a serious lack of tools for
analyzing many-body dynamics, in the classical case and even more so in the
quantum case.

In this lecture, we will discuss the derivation of the Euler equations from
microscopic quantum dynamics.  As we want to consider the genuine quantum
dynamics for a system with short-range pair interactions, we cannot take a
semiclassical limit. Although one-particle quantum dynamics converges to
Newtonian dynamics in the limit of infinite mass, this is not the case in the
thermodynamic limit, i.e., the heavy-particle limit does not commute with the
infinite-number-of-particles limit. This is most clearly seen in the pressure
function, for which quantum corrections survive at the macroscopic scale. In
fact, one of the conclusions of our work is that under rather general
conditions, the pressure function is the only place where the quantum nature of
the underlying system, in particular the particle statistics, survives in the
Euler limit. At the same time, our derivations also shows that it is the
quantum mechanical pressure, without modification, which governs the
macroscopic dynamics. A similar result should hold for all systems of
macroscopic conservation laws.

The Euler equations have traditionally been derived from  the Boltzmann
equation both in the classical case and in the quantum case, see Kadanoff and
Baym \cite{KB} for the quantum case.  Since the Boltzmann equation is valid
only in  very low density regions, these derivations  are not  satisfactory,
especially in the quantum case where the relationship  between  the quantum
dynamics and the Boltzmann equation is not entirely clear. There were, however,
two  approaches based directly on quantum dynamics. The first was due to Born
and Green \cite{BG}, who used an early version of what was later called the
BBGKY hierarchy, together with mo\-ment methods and some truncation
assumptions.  A bit later, Irving and Zwanzig \cite{IZ} used the Wigner
equation, moment methods and truncations to accomplish a similar result.  These
two approaches rely essentially on the moment method with the Boltzmann
equation replaced by the Schr\"odinger equation.  Unlike in the Boltzmann case,
where one can do asymptotic analysis to justify this approach, it seems
unlikely that this can be done for the Schr\"odinger dynamics. Therefore,
in the present work, we follow a much more direct route to connect the micro-
and macroscopic dynamics.

\section{Schr\"{o}dinger and Euler dynamics}\setzero
\vskip-5mm \hspace{5mm}

We begin by considering $N$ particles on $\RR^3$, evolving according
to the Schr\"o\-dinger equation
$$
        i \partial_t \psi_{t}(x_1, \cdots, x_N)
        =    H  \psi_{t}(x_1, \cdots, x_N)
$$
where the Hamiltonian is given by
\begin{equation}
        H = \; \;
 \sum_{j=1}^N  \frac { - \Delta_{j}}{ 2} +
\sum_{1\le i < j \le N}
        W (x_i-x_j) \; .
\label{Ham}
\end{equation}
Here, $W$ is a two-body short-ranged super-stable isotropic pair interaction
and  $\psi_{t}(x_1, \cdots, x_N)$ is the wave function of
particles  at time $t$. We consider spinless Fermions and thus the state space
${\cal H}^N$ is the subspace of antisymmetric functions in $L^2(\RR^{3N})$.
It is convenient not to fix the total number of particles
and to use the second quantization terminology. In fact, it would be
extremely cumbersome to work through all arguments without the second
quantization formalism.
The state space of the particles, called the Fermion Fock space,
is thus the direct sum of ${\cal H}^N$: $
 {\cal H} : =   \oplus_{N=0}^\infty \;
{\cal H}^{N}$.

Define the annihilation  and creation
operators $a_x$ and $a_x^+$ by
\beann
(a_x \Psi)^{N} ( x_1, \cdots, x_N)
& = & \sqrt {N+1} \Psi^{N+1}
(x, x_1, \cdots, x_N), \\
(a_x^+ \Psi)^{N} ( x_1, \cdots, x_N) & = & \frac {1 }{\sqrt {N}}
\sum_{j=1}^N  (-1)^{j-1} \delta(x-x_j) \Psi^{N-1} (x_1, \cdots,
\widehat {x_j}, \cdots, x_N),
\eeann
 where $a_x$ and $a_x^+$ are
operator-valued distributions and, as usual, $\hat{\; }$ means
``omit''.

The  annihilation  operator  $a_x$ is the
adjoint of $a_x^+$ with respect to the
standard inner product of the Fock space with Lebesgue measure $dx$, and
$$
 [ a_x, a^+_y ]_+ : = a_x a^+_y  + a_x a^+_y = \delta (x-y) \; ,
$$
where $\delta$ is the delta distribution. The derivatives of these distributions
with respect to the parameter $x$ are denoted by $\nabla a_x$ and
$\nabla a^+_x$. With this notation, we can express
the Hamiltonian as $H= H_0 + V$ where
$H_0 = \frac 1 2 \int \nabla a_x^+  \nabla a_x \, d x$ and
$V = \frac 1 2 \int \int dx dy W(x-y)
a_{x}^+ a_{y}^+ a_{y} a_{x}$.
It is more convenient to put the Schr\"odinger equation into the operator form,
which is sometimes called the Schr\"odinger-Liouville equation. Denote the
density matrix of the state at time $t$ by $\g_t$. Only normal states, which can
be represented by density matrices, will be considered in the time evolution.
Then the Schr\"odinger equation is equivalent to
$i\partial_t \g_t = {\delta}_H  \gamma_t$,
with ${\delta}_H \gamma_t : = [H , \g_t]$ .
The  conserved quantities of the dynamics are
the  number of particles ,   the three
components of the  momentum and the energy. The local densities
of these quantities are denoted
by $\bu
=(u^\mu), \mu=0, \cdots, 4$, and are given by the following expressions:
\bea
u^0_x = n_x &=& a_x^+ a_x,  \nonumber\\
u^j_x= p_x^j &=& -\frac i 2   [ \nabla_j a^+_x  a_x- a^+_x
\nabla_j a_x  ], \qquad j=1, 2, 3,  \label{u-def}\\
u^4_x= h_x &=& \frac 1 2 \nabla a^+_x \nabla a_x + \frac 1 2 \int dy W(x-y) a_{x}^+ a_{y}^+ a_{y} a_{x}. \nonumber
\eea The finite volumes, denoted by $\Lambda$, will always by three-dimensional tori and, unless otherwise stated,
unbounded observables on $\Lambda$ will be defined with periodic boundary conditions. E.g., the number of
particles in $\Lambda$, the total momentum, and the total energy of the particles in $\Lambda$, respectively, are
defined by
$$
N_\Lambda = \int_\Lambda dx\, n_x ,\quad P^j_\Lambda =
\int_\Lambda dx\, p^j_x,\quad j=1, 2, 3,\quad H_\Lambda =
\int_\Lambda dx\, h_x \,.
$$

We slightly generalize the definition of the grand canonical Gibbs
states to include a parameter for the total momentum of the
system: the Lagrange multiplier $\alpha\in\Rl^3$. We will work
under the assumption that the temperature and chemical potential
are in the one-phase region of the phase diagram of the system
under consideration, such that the thermodynamic limit is unique.
The infinite volume Gibbs states are then given by the following
formula: \be \omega_{\beta, \alpha, \mu}(X)=\lim_{\Lambda\to\RR^d}
\frac{\Tr Xe^{ -\beta( H_{0,\Lambda}+V_\Lambda-\alpha \cdot
P_\Lambda- \mu N_\Lambda)}}{\Tr
e^{-\beta(H_{0,\Lambda}+V_\Lambda-\alpha P_\Lambda- \mu
N_\Lambda)}}\,.
\end{equation}
It is convenient to denote the parameters
$(\beta, \alpha, \mu)$ by $\vec \lambda= (\lambda^\mu), \mu=0, \cdots, 4$
with $\lambda^0= \beta \mu, \lambda^j = \beta \alpha^j, \lambda^4 = \beta
$. Define (notice the sign convention)
$$
 {\vec \lambda}\cdot  \vec  u
= \sum_{\mu=0}^3  { \lambda}^\mu \,   u  ^\mu\, - \, { \lambda}^4
\,   u  ^4\quad \mbox{and}\quad \lan  {\vec \lambda},  \vec  u
\ran_\Lambda = |\Lambda|^{-1} \int_\Lambda  d x   {\vec
\lambda}(x) \cdot  \vec  u  (x)\,.
$$
These notations allow us to give a compact formula for the unique,
translation invariant Gibbs state (defined with constant $\vec
\lambda$), as well as for the states describing local equilibrium
(defined with $x$-dependent $\vec \lambda$): \be \omega_{\vec
\lambda} = \lim_{\Lambda\to\RR^d} e^{ |\Lambda| \lan {\vec
\lambda}, {\vec  u  } \ran_\Lambda }/Z_{\Lambda} ({\vec \lambda} )
\label{omega-def}\ee where $Z_{\Lambda}({\vec \lambda} ) $ is the
partition function given by $ Z_{\Lambda} ({\vec \lambda} ) = \Tr
e^{\vert\Lambda\vert \lan {\vec \lambda}, {\vec  u  } \ran_\Lambda
}$. If we define the pressure, as a function of the constant
vector $\vec \lambda$, by $\psi ({\vec \lambda} ) =  \lim_{L \to
\infty} |\Lambda|^{-1} \log Z_{\Lambda} ({\vec \lambda} )$, then
\be \frac {\partial \psi } {\partial \lambda^\mu} = \omega_{\vec
\lambda}(  u  ^\mu ).
 \label{dual}\ee
 As the states
$\omega_\lambda$ are translation invariant, we have
$\omega_\lambda(u^\mu)=q^\mu$. Explicitly,
$$
\rho =\omega_\lambda(n_x),\quad
%=\lim_{\Lambda\to\RR^3}\frac{1}{\vert\Lambda\vert}
%\omega_\lambda(N_\Lambda)\\
\fq =\omega_\lambda(p_x),\quad
%=\lim_{\Lambda\to\RR^3}\frac{1}{\vert\Lambda\vert}
%\omega_\lambda(P_\Lambda)\\
e=\omega_\lambda(h_x).
%=\lim_{\Lambda\to\RR^3}\frac{1}{\vert\Lambda\vert}\omega_\lambda(H_\Lambda)
$$
Notice that $\fq$ and $e$ are momentum and energy per volume.

Again, we will work under the assumption that these parameters stay in the
one-phase region, the limiting Gibbs state is unique and these definitions are
unambiguous. Although momentum is preferable as a quantum observable, we also
introduce the velocity in order to be able to compare with the classical case.
The velocity field $\fv(x)$ has to be defined as a mean velocity of the
particles in a neighborhood of $x$. Therefore we have $\fv(x)=\fq(x)/\rho(x)$.
We also introduce the energy per particle defined by $\tilde e  = e /\rho$. The
usual Euler equations are written in terms of $\rho, \fv$, and $\tilde e $.

In order to derive the Euler equations, we need to perform a
rescaling. So we shall put all particles in a torus $\Lambda_\e$ of
size $\e^{-1}$ and use $(X,T) = (\e x, \e t)$ to denote the
macroscopic coordinates. For all equations in this paper periodic
boundary conditions are implicitly understood.

The Euler equations are given by
\bea
\frac{\partial \rho}{\partial T} +\sum_{j=1}^3
\frac{\partial}{\partial
X_j} (\rho \fv_j)  &=& 0, \nonumber \\
\frac{\partial \, (\, \rho \fv_k \, ) }{\partial T}
+\sum_{j=1}^3 \frac{\partial}{\partial
X_j} \left [ \; \rho \fv_j  \fv_k  \; \right ]
+\frac{\partial}{\partial X_k}P(e , \rho)
&=& 0, \label{euler}\\
\frac{\partial \, (\, \rho \tilde e  \, ) }{\partial T} + \sum_{j=1}^3
\frac{\partial}{\partial
X_j}\left [ \; \rho \tilde e   \fv_j  + \fv_j P(e , \rho) \; \right ]
&=& 0.  \nonumber %\label{euler3}
\eea
These equations are in form identical to the classical ones but all
physical quantities are computed quantum mechanically.
In particular, $P(e , \rho)$ is the thermodynamic pressure computed from
quantum statistical mechanics for the microscopic system. It is a function
of $X$ and $T$ only through its dependence on $e$ and $\rho$. If no
velocity dependent forces act between the molecules of the fluid under
consideration (we consider only a pair potential), the pressure is
independent of the velocity.

Let $\vec q= (q^0, \cdots, q^4)$, related to density, momenta and energy as
follows:
\be
q^0 = \rho \ , \quad q^i   = \rho \, \fv^i \ , \quad q^4 =e=
 \rho \, \tilde e.
\label{2.17} \ee In other words $q^1, q^2, q^3,$ and $q^4$ are
momenta and energy per {\it volume} instead of {\it per particle}
as in the usual Euler equation \eq{euler}. We rewrite the Euler
equations in the following  form \be \frac {\partial q^k }{
\partial T} +\sum_{i=1}^3 \nabla_i^X \big[ A^k_i (q)\big] = 0 \ ,
\quad  k =0,1,2,3,4. \label{2.15}
\end{equation}
The matrix $A$ is determined by comparison with the Euler equations:
$$
A_j^0 =q^j,\
A^i_j =\delta_{ij} P + q_i q_j /q _0,\
A^4_j = q^j (q_4 +P)/q_0.
$$

\section{Local equilibrium}\setzero
\vskip-5mm \hspace{5mm}

To proceed we need a microscopic description of local
equilibrium. Suppose we are given macroscopic
functions $\vec q (X)$. We wish to find a local Gibbs state
with the conserved quantities given by $\vec q (X)$.
The local Gibbs states are states locally in equilibrium.
In other words,  in a microscopic
neighborhood of any point $x\in T^3$ the state is given by
a Gibbs state. More precisely, we wish to find
a local Gibbs state with
the expected values of the energy, momentum, and particle number per
unit volume at $X$ given by $\vec q (X)$. To achieve this, we
only have to adjust the parameter $\vec \lambda$ at every point $X$.
More precisely, we choose ${\vec \lambda} (X)$ such that the equation
\eq{dual} holds at every point, i.e.,
$$
\frac {\partial \Psi ({\vec \lambda} (X))} {\partial
{ \lambda}^\mu (X)} =
 q^\mu (X) .
$$
If we denote the solution to the Euler equation
by $q(X, T)$, then we can choose in a similar way
a local Gibbs state with given conserved quantities at the time $T$.
Define
%$\vec \lambda_\e (t, x) = \vec \lambda(\e t, \e x)$  and define
the local Gibbs state
\be
\omega_t^\e = \frac {1}{c_\e (t)}\exp  \left [\,\e^{-3}  \lan
 {\vec \lambda}(\e t,\e \,\cdot\,), \, \vec  u   \ran_{{\e^{-1} }} \right ]
\label{2.27}
\end{equation}
where $c_\e (t)$ is the normalization constant.
Clearly, we have that $\omega_\e (t) ( u  _x^\mu ) = q^\mu (\e x, \e t)$
to leading order in $\e$.

In summary, the goal is to show that, in the limit $\epsilon\to0$, the following
diagram commutes:
$$\begin{CD}
\bq(X,0)@>\mbox{Euler}>>\bq(X,T)\\
@V\mbox{local equilibrium}VV @AA\parbox{4cm}{limit $\epsilon\downarrow 0$
of expectation of locally averaged observables}A\\
\gamma_0@>\mbox{Schr\"odinger}>>\gamma_{\epsilon^{-1}T}
\end{CD}$$

As smooth solutions of the Euler equations are guaranteed to exist only up to a
finite time \cite{KM}, say $T_0$, we will formulate our assumptions on the
dynamics of the microscopic system for a finite time interval as well, say
$t\in [0, T_0/\e]$. Note the cutoff assumptions below would hold automatically
for lattice models.

\section{Assumptions and the main theorem}\setzero
\vskip-5mm \hspace{5mm}

Our main result is stated in Theorem \ref{thm-1} below. First, we state the
assumptions of the theorem with some brief comments. There are three kinds
of assumptions.

The first category of assumptions could be called {\em physical\/} assumptions
on the solution of the Euler equations that we would like to obtain as a
scaling limit of the underlying dynamics, and on the pair interaction potential
of this system.

\smallskip \noindent {\bf I. One-phase regime:} We assume that the
pair potential, $W$, is $C^1$ and with support contained in a ball
of radius $R$. Moreover, we assume that $W$ is  symmetric under
reflections of each of the coordinate axes (e.g., rotation
symmetric potentials automatically have this symmetry), and has
the usual stability property \cite{Rue}: there is a constant
$B\geq 0$ such that,  for all $N\geq2$, $x_1,\ldots,x_N\in\Rl^3$,
$$ \sum_{1\leq1<j\leq N}W(x_i-x_j)\geq -BN. $$

Of the Fermion system with potential $W$ we assume that there is an open region
$D\, \subset \, \Rl^2$, which we will call the {\em one-phase region\/}, such that the
system has a unique limiting Gibbs state for all values of particle density and
energy density  $(\rho,e)\in D$.

The solution of the Euler equations we consider, $q(X,T)$, will be assumed to
$C^1$ in $X$ for $T\in [0,T_0]$, and have local particle and energy density
in the one-phase region for all times $T\in [0,T_0]$. I.e., $(\rho(X,T),
e(X,T))\in D$, for all $X\in \Lambda_1$ and $T\in [0,T_0]$.

The next category of assumptions is on the local equilibrium states for the
Fermion system that we construct and on their time-evolution under the
Schr\"{o}dinger equation. They can be considered conjectures. In fact, these
assumptions have not been rigorously proved even for Gibbs states. Although one
expects that these assumptions can be proved in the high-temperature and
low-density region using some
type of cluster expansion methods, this has only been done recently for Bosons
in \cite{GLM}. For the rest of this paper, we shall assume this cutoff assumptions
for the solution to the Schrodinger equation that we consider,  as well as for
the Gibbs states in the one phase regions considered in this paper.

\noindent {\bf II. Cutoff assumptions:} Suppose that $\gamma_t$ is
the solution to the Schr\"odinger equation with a local
equilibrium state as initial condition, constructed with the
parameters derived from a solution of the Euler equations (with
the appropriate pressure function) that does not leave the
one-phase region.  We make the following two assumptions about
this solution:

\noindent {\em 1. Finite velocity cutoff assumption:} Let $N_p(t)= \Tr \gamma_t  a_p^+  a_p$, where $a^\#_p$ is
the Fourier transform of $a^\#_x$. Then there is a constant $c > 0$, such that for all $t \le T_0/\e$,
$$
\e^d  \int dp   e^{ c p^2} N_p(t) \le C_{T_0}.
$$

\noindent {\em 2. Non-implosion assumption:} There is a constant $C_{T_0}$, such that for all $t \le T_0/\e$, \be
%\frac \e T   \int_0^{T /\e} dt
\Tr \gamma_t  \e^{d} \int dx n_x \left [  \int_{|x-y| \le 2 R}   n_y  d y \right ]^2 \le C_{T_0}, \label{2bound}
\end{equation}
where $R$ is the range of the   interaction $W$.

Finally, we have an assumption on the set of the time-invariant ergodic states
of the Fermion system. To state this assumption we need the notion of
{\em relative entropy\/}, of a normal state $\gamma$ with respect to another
normal state $\omega$. Let $\gamma$ and $\omega$ denote the density matrices
of these states. The relative entropy, $S(\gamma\mid\omega)$, is defined by
$$
S(\gamma|\omega)= \begin{cases} \Tr \gamma(\log \gamma -\log
\omega) & \mbox{if } \ker \omega  \subset  \ker \gamma,\\
+\infty &\mbox{else}.
\end{cases}
$$
For a pair of translation invariant locally normal states, one
can show existence of the relative entropy density \cite{OP},
defined by the limit
$$
s(\gamma|\omega)= \lim_{\e\downarrow 0}\e^3
S(\gamma_{\Lambda_\e}\mid\omega_{\Lambda_\e})\; ,
$$
where $\gamma_{\Lambda_\e}$ and $\omega_{\Lambda_\e}$ denote the
density matrices of the normal states obtained by restricting $\gamma$
and $\omega$ to the observables localized in $\Lambda_\e=\e^{-1}\Lambda_1$.
The existence of the limit can be proved under more general conditions
on the finite volumes, but this is unimportant for us.

\noindent {\bf III. Ergodicity assumption (``Boltzmann
Hypothesis''):} All translation invariant ergodic stationary
states to the Schr\"odinger equation are Gibbs states if they
satisfy the following assumptions: {\em 1) the density and energy
is in one  phase region}. {\em 2) The  relative entropy density
with respect to some Gibbs state is finite}.

We expect that our assumptions hold for the solutions $\gamma_t$ of the
Schr\"odinger equation that we employ, but it must be said that, at this
moment, very little is known. We believe that these  are natural conditions. To
prove them under rather general conditions or even for a special class of
models may be regarded as an important open problem in quantum statistical
mechanics.

\begin{theorem} \label{thm-1}
Suppose that $\vec q(X, T)$ is a smooth solution to the Euler equation
in one phase region
up to time $T\le T_0$. Let $\omega^\e_t$ be the local Gibbs state
with conserved quantities given by $\vec q(X, T)$.
Suppose that the  cutoff assumptions and the ergodicity assumption hold.
Let $\gamma_t$ be the solution to the Schr\"odinger equation
 and $\gamma_0 =\omega^\e_0$ (Notice  $\gamma_t$ depends on $\e$).
Then for all $t \le T_o/\e$ we have
$$
\lim_{\e \to 0} s(\gamma_t|\omega^\e_t) = 0.
$$
In other words, $\omega^\e_t$ is a solution to the Schr\"odinger equation
 in entropy sense.
In particular, for any smooth function $f$ on $\Lambda$, we have
$$
\lim_{\e \to 0} \e^3\int_{\Lambda_\e} d x f(\e x)\left [ \gamma_t
(\vec  u  _x) - \vec  q(\e t, \e x) \right ] = 0.
$$
\end{theorem}

This theorem is a quantum analogue of the classical result by Olla, Varadhan,
and Yau \cite{OVY}. There are a few differences in the assumptions and the
strategy followed in their paper with respect to ours. E.g.,
in the treatment of \cite{OVY} of the classical case the cut-off assumption 1)
was not needed. Instead, the usual quadratic kinetic energy was replaced by one
with bounded derivatives with respect to momentum. For Fermion models on a
lattice instead of in the continuum, no cut-off assumptions are required.
Another difference with the treatment in \cite{OVY}, is
that we do not add noise terms to the ``native'' dynamics. In \cite{OVY} a weak
noise term was added to the Newtonian dynamics in order to be able to prove
convergence  to local equilibrium. In that paper, the strength of the noise
vanishes in the hydrodynamic limit. Here, we do not modify the Heisenberg
dynamics in any way, but instead reduce the question of convergence to local
equilibrium to the ergodicity property given in Assumption III.

\section{Outline of the proof}\setzero
\vskip-5mm \hspace{5mm}

The basic structure of our proof follows the relative entropy
approach of \cite{OVY, Yau}. The aim is to derive a differential
inequality for the relative entropy between  the solution to the
Schr\"odinger equation and a time-dependent local Gibbs state
constructed to reproduce the solution of the Euler equations. The
time derivative of the relative entropy can be expressed as an
expectation of the local currents with respect to the solution to
the Schr\"odinger equation: Since  $\gamma_t$ is a solution to the
Schr\"odinger equation, we have  for any density matrix
$\omega^\e_t$ the identity \be \frac d {dt}
S(\gamma_t|\omega^\e_t) = \Tr \gamma_t \left \{   i \adH  -
\partial_t  \right \} \log \omega^\e_t. \label{re}
\end{equation}
Using the definition of the local equilibrium states \eq{omega-def}, we get
\be
\frac d {dt} s(\gamma_t|\omega^\e_t)
=    \e^3  \Tr \gamma_t \left \{   i \adH  - \partial_t   \right \}
\left \{
 \lan  {\vec \lambda_\e } (t, \cdot ) ,   \bu \ran
 - \log c_\e (t) \right \}.
\label{re2}
\end{equation}
Direct computation yields that
$  \adH  \lan
 {\vec \lambda}_\e  ( t, \cdot),   \bu \ran
=  - \e \sum_{j=1}^3   \lan  \nabla_j {\vec \lambda_\e } (t, \cdot ), \;
{\vec \theta}_{j}(t) \ran
$, to leading order in $\e$, and where the ${\vec \theta}_{j}$
are local observables for the microscopic currents of the conserved
quantities.
Since we do not know the solution of the Schr\"odinger equation, $\gamma_t$,
well enough, this expectation in the RHS of \eq{re2} cannot be computed or
estimated explicitly. The main idea is to bound these expectations in terms of
the relative entropy itself, thus obtaining a differential inequality. All
these bounds are based on the following
inequality, which is a consequence of the variational principle for
the relative entropy \cite{Pet,OP}:
for density matrices/states $\omega$ and $\gamma$, with
$\ker\omega=\{0\}$, and for all self-adjoint $h$, and $\delta > 0$, one has
\begin{equation}
\gamma(h) \le \delta^{-1} \log\Tr \; e^{ \delta h+ \log \omega }
+\delta^{-1} S(\gamma|\omega). \label{entineq}
\end{equation}
Most of the work is then to show that the terms resulting from the first term
in the RHS of this inequality are sufficiently small. This requires a number
of steps. Two essential ingredients are the Euler equations (naturally), and
a quantum version of Virial Theorem to relate certain terms in the currents
to the thermodynamic pressure as given by quantum statistical mechanics.
Next, we explain the main steps in a bit more detail.

{\sl Step 1: Replace the local microscopic currents by macroscopic currents.}
The basic idea in hydrodynamical limit is first to show that the local space
time average of the solution is time invariant. From Assumption III, ergodic
time invariant states are Gibbs. For Gibbs states, we can replace the local
microscopic currents by macroscopic currents.

{\sl 1a: Construct a commuting version of local conserved quantities.}
In principle, one should be able to express the macroscopic currents as
functions of the conserved quantities. The local conservative quantities,
however, are operators which commute only up to boundary terms. Therefore,
we either need to prove that the non-commutativity does not affect the
meaning of macroscopic currents or we need to construct some commuting version
of the local conservative quantities. We follow the second approach and
construct a commuting version of local conservative quantities with a method
inspired by \cite{CLY}.

{\sl 1b: Restriction to the one phase region.}
Since Assumption III can only be applied in the one phase region,
we have to introduce suitable cutoff functions. As these,
in general, do not commute with the local currents, this
is a non-trivial step.

{\sl 1c: Apply the Virial Theorem.}
Following the method of \cite{NV}, we prove and apply a quantum virial theorem
to relate the local currents to the pressure.

{\sl Step 2: Estimate all errors by local conservative quantities.}
For this step the entropy inequality \eq{entineq} is crucial.

{\sl Step 3: Derive a differential inequality of the  entropy with error term
given by a large deviation formula. } As there is no large deviation theory for
non-commuting observables, it is essential that we have expressed everything by
commuting objects.

The rest of the argument follows by the standard relative entropy method.
The main technical difficulties, in comparison with the classical
case, all stem from the non-commutativity of the algebra of
observables. Simple inequalities, such as $|A+B| \le |A|+|B|$
and $\vert AB\vert\le \vert A\vert\,\vert B\vert$, which
are used numerous times in estimates for classical systems, are false
for quantum observables. Therefore, all estimates have to be derived
without taking absolute values. A full account of our work will appear
shortly \cite{NY}.

\noindent{\bf Acknowledgments.} This material is based upon work
supported by the National Science Foundation under Grants \#
DMS-0070774 (B. N.) and \# DMS-9703752 (H.-T. Y.)

\label{lastpage}

\end{document}